\documentclass[a4paper]{jpconf}
\usepackage{graphicx,amsmath}
\usepackage{slashed}
\usepackage{amssymb}

\begin{document}

\newcommand{\be}{\begin{equation}}
\newcommand{\ee}{\end{equation}}
\newcommand{\beq}{\begin{eqnarray}}
\newcommand{\eeq}{\end{eqnarray}}

\title{Aspects of the Bosonic Spectral Action}

\author{Mairi Sakellariadou}

\address{Physics Department, King's College London, University of London\\
Strand, London WC2R 2LS, UK}

\ead{Mairi.Sakellariadou@kcl.ac.uk}

\begin{abstract}
  A brief description of the elements of noncommutative spectral
  geometry as an approach to unification is presented. The physical
  implications of the doubling of the algebra are discussed. Some high
  energy phenomenological as well as various cosmological consequences
  are presented. A constraint in one of the three free parameters,
  namely the one related to the coupling constants at unification, is
  obtained, and the possible r\^ole of scalar fields is highlighted. A
  novel spectral action approach based upon zeta function
  regularisation, in order to address some of the issues of the
  traditional bosonic spectral action based on a cutoff function and a
  cutoff scale, is discussed.
\end{abstract}

\section{Introduction}
The Standard Model (SM) of strong and electroweak interactions remains
the most successful particle physics model we have at hand and its
validity has been recently confirmed by the discovery of the Higgs
boson. However, several conceptual questions remain unanswered, while
it may be necessary to go beyond the SM, possibly relating it to a
theory of Quantum Gravity. Noncommutative Spectral Geometry (NCSG)
aims at explaining some of the conceptual issues of the SM, whilst it
offers a new geometrical framework to address physics at the Quantum
Gravity regime. To construct a Quantum theory of Gravity coupled to
matter, one may either neglect matter altogether (as for instance
within the framework of Loop Quantum Gravity), or consider instead
that the interaction between gravity and matter is the most important
ingredient to define the dynamics.  Noncommutative Spectral
Geometry~\cite{ncg-book1,ncg-book2} follows the latter approach,
aiming at defining the noncommutative algebra of observables of a
Quantum theory of Gravity.

Noncommutative spectral geometry starts from the following remark. At
energy scales much below the Planck scale it is reasonable to assume
that physics can be described by continuum fields and an effective
action (the sum of the Einstein-Hilbert and SM actions), but close to
Planck scale this assumption is no longer valid and Quantum Gravity
effects may imply that space-time is a heavily noncommutative
manifold. Remaining close but below Planck scale, one can however
consider that the algebra of coordinates is only a lightly
noncommutative algebra of matrix valued functions, and by chosen
properly this algebra NCSG leads to a purely geometric explanation of
the SM coupled to gravity~\cite{ccm}. In the context of NCSG, gravity
and the SM fields are packaged into geometry and matter on a
Kaluza-Klein noncommutative space and using well-established
experimental results at the electroweak scale, we can guess the
small-scale space-time structure avoiding an {\sl ad hoc} proposal. In
that sense, noncommutative geometry can be considered as a bottom-up
approach, complimentary to the top-down string theory approach.

Noncommutative spectral geometry proposes to consider the SM as a
phenomenological model which dictates the space-time geometry.  In
this way, the geometric space is defined as the product of a
4-dimensional compact Riemannian manifold ${\cal M}$, describing the
geometry of space-time, with an internal zero-dimensional discrete
finite internal Kaluza-Klein space ${\cal F}$, describing the internal
geometry, attached to each point. Such simple noncommutative spaces
${\cal M}\times {\cal F}$, where the noncommutative algebra describing
space is the algebra of functions over ordinary space-time, are called
{\sl almost commutative manifolds}. Note that such spaces are
different from the more general noncommutative spaces such as the
Moyal plane for which $[x^i, x^j]=i\theta^{ij}$, where $\theta^{ij}$
is an anti-symmetric real $d\times d$ matrix (with $d$ the space-time
dimensionality) representing the fuzziness of space-time.  In NCSG,
the description of ordinary Riemannian manifolds in terms of spectral
data, is extended for noncommutative manifolds. Hence, one defines the
almost commutative manifold ${\cal M}\times{\cal F}$ by a spectral
tripe, and dynamics are given by a spectral action that sums up all
frequencies of vibration of space. At last, one aims at answering
whether we can hear the shape of such a spectral triple, called a {\sl
  spinorial drum}.

In the following, we will give a short description of the main elements of
NCSG~\cite{Walterbook}-\cite{Sakellariadou:2012jz} , discuss its
phenomenological particle physics consequences, the physical
implications of the precise construction of the almost commutative
manifold~\cite{PRD,mvpm}, and examine the cosmological consequences of
the gravitational sector of the
theory~\cite{Nelson:2008uy}-\cite{sgaam}. We will then address
some issues regarding the traditional bosonic spectral action approach
and highlight a new proposal~\cite{mfma}.

\section{Elements of Noncommutative Spectral Geometry}
Let us start with the case of spin (in order to be able to describe
spinors) manifolds in NCSG. Given a compact 4-dimensional Riemannian
spin manifold ${\cal M}$, consider the set $C^\infty({\cal M})$
of smooth infinitely differentiable functions, and the Hilbert space
${\cal H}=L^2({\cal M}, S)$ of square-integrable spinors $S$ on ${\cal
  M}$. One can show that the set $C^\infty({\cal M})$ is an algebra
${\cal A}$ under point-wise multiplication, acting on ${\cal H}$ as
multiplication operators. Let us also consider $\slashed {\cal D}$,
the Dirac operator $-i\gamma^\mu\nabla_\mu^s$, acting as first order
differential operator on the spinors. The algebra, Hilbert space and
Dirac operator form the canonical triple $(C^\infty({\cal M}),
L^2({\cal M}, S), \slashed {\cal D})$. In addition, we consider the
$\gamma_5$ operator with $\gamma_5^2=1, \gamma_5^\star =\gamma_5$,
which is just a $ \mathbb{Z}_2$-grading, that decomposes the Hilbert
space ${\cal H}$ into a positive and negative eigenspace $L^2({\cal
  M}, S)=L^2({\cal M}, S)^+ \oplus L^2({\cal M}, S)^-$, hence playing
the r\^ole of a chirality operator. We also consider an antilinear
isomorphism $J_{\cal M}$ with $J_{\cal M}^2=-1, J_{\cal M}\slashed
{\cal D}=\slashed {\cal D}J_{\cal M}, J_{\cal M}\gamma_5=\gamma_5
J_{\cal M}$, as the charge conjugation operator on spinors.

Consider now an almost-commutative manifold ${\cal M}\times {\cal
  F}$. The canonical triple defining ${\cal M}$ encodes the space-time
structure, whereas the triple $({\cal A}_{\cal F}, {\cal H}_{\cal F},
D_{\cal F})$ encodes the internal degrees of freedom at each point in
space-time, allowing a description of a gauge theory on the spin
manifold ${\cal M}$. To obtain the SM, the most important ingredient
is the choice of the matrix algebra ${\cal A}_{\cal F}$, acting on the
Hilbert space ${\cal H}_{\cal F}$ via matrix multiplication. The
operator ${\cal D}_{\cal F}$ is a $96\times 96$ matrix expressed in
terms of the $3\times 3$ Yukawa mixing matrices and a real constant
responsible for the neutrino mass terms. This operator corresponds to
the inverse of the Euclidean propagator of fermions.  In addition, we
consider a $\gamma_{\cal F}$ grading such that $\gamma_{\cal F}=+1$
for left-handed fermions and $\gamma_{\cal F}=-1$ for right-handed
ones, and a conjugation operator $J_{\cal F}$ for the finite space
${\cal F}$. The almost-commutative manifold ${\cal M}\times{\cal F}$
is expressed by the spectral triple $({\cal A, H}, D)$:
\begin{equation}
{\cal M}\times{\cal F}:=(C^\infty({\cal M},{\cal A_F}), L^2({\cal M},
S)\otimes {\cal H_F}, \slashed {\cal
  D}\otimes\mathbb{I}+\gamma_5\otimes {\cal D}_{\cal F})~.  \nonumber
\end{equation}
The choice of the algebra ${\cal A_F}$ is the most important input of
the NCSG approach to the SM, and has to be chosen appropriately. For
instance, ${\cal A_F}$ cannot be right-handed symmetric. It has been
shown~\cite{Chamseddine:2007ia} that this algebra has to be the
product of the algebra of quaternions and the algebra of the complex
$k\times k$ matrices with $k$ an even number $k=2a$:
\begin{equation}
{\cal A_F}=M_a(\mathbb{H})\oplus M_k(\mathbb{C})~.
\nonumber
\end{equation}
The first value of $k$ that produces the correct number of fermions
(namely 16) in each of the three generations, is $k=4 $. Hence, NCSG
predicts that the number of fermions is the square of an even integer,
while the existence of three generations is just a physical
input. Note that the particular choice of Hilbert space is of no
importance, since all separable infinite-dimensional Hilbert spaces
are isomorphic.  Hence, the fermions of the SM provide the Hilbert
space of a spectral triple for the algebra, while the boson of the SM,
including the Higgs boson, are obtained through inner fluctuations of
the Dirac operator of the product ${\cal M}\times {\cal F}$ geometry.
Thus, the Higgs boson becomes just a gauge field corresponding to a
finite difference.

To derive a physical Lagrangian one then applies the spectral action
principle, stating that the action functional depends only on the
spectrum of the fluctuated Dirac operator ${\cal D}_A$:
\begin{equation}
{\cal D}_A={\cal D} + A + \epsilon' J A J^{-1}~,
\end{equation}
with $A=A^\star$ a self-adjoint operator of the form
\begin{equation}
A=\sum_j a_j[{\cal D}, b_j]~;~a_j,b_j\in {\cal A}~,
\end{equation}
$J$ an anti-unitary operator such that $J^2=1$ and $\epsilon\in\{\pm 1\}$,
and is of the form
\begin{equation}
{\rm Tr}(f({\cal D}^2_A/\Lambda^2))~,
\label{bosonic}
\end{equation}
with $f$ a cut-off function and $\Lambda$ denoting the energy scale at
which this Lagrangian is valid. More precisely, $f$ is a positive
function that falls to zero at large values of its argument, so that
the integrals $\int_0^\infty f(u) u du$ and $\int_0^\infty f(u) du$
are finite. Typical cut-off functions $f$ used in the literature are
$f(x)=1$ for $x\leq \Lambda$, or $f(x)=e^{-x}$.  The action given in
Eq.~\eqref{bosonic} above, sums up all eigenvalues of the fluctuated
Dirac operator ${\cal D}_A$ which are smaller than the cut-off energy
scale $\Lambda$. This trace can be then evaluated using heat kernel
techniques and thus expressed through the Seeley-de Witt coefficients
$a_n$, known for any second order elliptic differential operator, as
$\sum_{n=0}^\infty F_{4-n}\Lambda^{4-n} a_n$ where $F$ is defined as
$f({\cal D}_A^2)$.

The spectral action can be expanded in powers of the scale $\Lambda$
in the form~\cite{ac1997}
\begin{equation}
\label{eq:sp-act}
{\rm Tr}\left(f\left(\frac{{\cal D}_A^2}{\Lambda^2}\right)\right)\sim 
\sum_{k\in {\rm DimSp}} f_{2k} 
\Lambda^{2k}{\int\!\!\!\!\!\!-} |{\cal D}_A|^{-2k} + f(0) \zeta_{{\cal D}_A^2}(0)+ {\cal O}(1)~,
\end{equation}
where $f_{2k}$ are the momenta of the function $f$, defined as
\be\label{eq:moments0}
f_{2k}\equiv\int_0^\infty f(u) u^{2k-1}{\rm d}u\ \ ,\ \ \mbox{for}\ \ k>0 ~,
\nonumber\\\ee
and $f_0\equiv f(0)$. The noncommutative integration is defined in
terms of residues of zeta functions, $\zeta_{{\cal D}_A^2} (s) = {\rm
  Tr}(|{\cal D}_A|^{-2s})$ at poles of the zeta function, and the sum
is over points in the {\sl dimension spectrum} of the spectral triple.

Since $f$ is a cut-off function, its
Taylor expansion vanishes at zero, implying that the asymptotic
expansion of the trace, namely
\begin{equation}
{\rm Tr}(f({\cal D}^2_A/\Lambda^2))\sim 2f_4\Lambda^4 a_0({\cal
  D}_A^2) + 2f_2\Lambda^2 a_2({\cal D}_A^2) + f(0)a_4({\cal D}_A^2)
+{\cal O}(\Lambda^{-2})~,
\end{equation}
can be only given from the three first terms of the expansion.
The cut-off function plays a r\^ole through only three of its momenta:
\begin{equation}
f_4=\int_0^\infty f(u)u^3du\ ;\ f_2=\int_0^\infty f(u)udu\ ; f_0=f(0)~.
\end{equation}
related to the cosmological constant, the gravitational constant and
the coupling constants at unification, respectively.  

The bosonic spectral action, Eq.~\eqref{bosonic}, must be seen {\sl
  \`a la} Wilson, hence as the bare action at the mass scale
$\Lambda$.  This action only accounts for the bosonic part. Hence, to
account for the terms involving fermions and their coupling to the
bosons, one needs to include the fermionic part, which for a
KO-dimension 2 almost commutative manifold reads
\begin{equation}
(1/2)\langle J\Psi, {\cal D}_A \Psi\rangle~;~ \Psi\in{\cal H}^+~.
\label{fermionic}
\end{equation}
After a long calculation one eventually obtains that the bosonic
spectral action at the cutoff scale $\Lambda$ and using the cutoff
normalisation through the cutoff function $f$, reads
\begin{eqnarray}
S_\Lambda ={-2a f_2\Lambda^2+e f_0\over \pi^2} \int
|\phi|^2\sqrt{g}d^4x +{f_0\over 2\pi^2}\int a|D_\mu\phi|^2\sqrt{g}d^4x
-{f_0\over 12\pi^2}\int aR|\phi|^2\sqrt{g}d^4x \nonumber\\ -{f_0\over
  2\pi^2}\int \left(g_3^2 G_\mu^i G^{\mu i}+g_2^2F_\mu^a F^{\mu\nu
  a}+{5\over 3}g_1^2B_\mu B^\mu\right) \sqrt{g}d^4x +{f_0\over
  2\pi^2}\int b|\phi|^4\sqrt{g}d^4x +{\cal O}(\Lambda^{-2})~,
\end{eqnarray}
with $a, b, c, d, e$ constants depending on the Yukawa parameters.
Adding to the above action the fermionic part, as indicated in
Eq.~(\ref{fermionic}), one obtains~\cite{ccm} the full SM Lagrangian.
Since this spectral action is characterised by the cutoff function $f$
and the cutoff scale $\Lambda$, we will call it the {\sl cutoff
  bosonic spectral action}, to differentiate it from another
regularisation procedure we will highlight later.

To discuss the particle physics phenomenological consequences of NCSG
let us briefly discuss the obtained Lagrangian. Its coefficients are
given in terms of the three momenta $f(0), f_2, f_4$ of the cut-off
function $f$, of the cut-off scale $\Lambda$, of the vacuum
expectation value of the Higgs field $\phi$, and of the coefficients
$a, b, c, d, e$, which are determined by the mass matrices in the
Dirac operator ${\cal D_F}$. Given that among the various relations
connecting the coefficients $a, b, c, d, e$, one finds
$g_2^2=g_3^2=(5/3)g_1^2$, which holds in several Grand Unified
Theories (GUTs) (like SU(5)), one may assume that the theory is valid
at the GUT scale. One then uses standard renormalisation group flow
techniques to obtain predictions for the SM phenomenology. For
instance, one finds that the top quark mass is $m_{\rm t}\leq 180$
GeV.  The NCSG approach leads to a Higgs doublet with a negative mass
term and a positive quartic term, hence implying the existence of a
spontaneously symmetry breaking mechanism of the electroweak
symmetry. Let us comment on the predicted value of the Higgs mass. The
NCSG model involves three scalars, namely a Higgs field, a singlet and
a dilaton. The singlet is a real scalar field associated with the
Majorana mass of the right-handed neutrino, having a nontrivially
mixing with the Higgs field.  In the original approach~\cite{ccm}, the
singlet was integrated out being replaced by its vacuum expectation
value, leading to an incorrect prediction of the Higgs mass, namely
$167 ~{\rm GeV}\leq m_{\rm h}\leq 176 ~{\rm GeV}.$ This conflict was
resolved in the subsequent approach~\cite{Resilience} where this
assumption was relaxed. Hence considering the mixing between the Higgs
doublet and singlet, consistency with the experimental result of a 125
GeV Higgs mass and a 170 GeV top quark mass was achieved. Note that
the r\^ole of the singlet field was already mentioned
previously~\cite{Stephan}. Moreover, the experimentally found Higgs
mass can be accommodated by either considering a model based on a
larger symmetry, the {\sl grand symmetry}, where the algebra is
$A_G=M_4({\mathbb H})\oplus M_8({\mathbb C})$ ~\cite{grand-algebra},
or by generalising the inner fluctuations to real spectral triples
that fail on the first order condition, leading to a Pati-Salam type
of model ${\rm SU}(2)_{\rm R}\times {\rm SU}(2)_{\rm L}\times {\rm
  SU}(4)$~\cite{Chamseddine:2013sia}.

Assuming the big desert hypothesis, one-loop renormalisation group
analysis for the three gauge couplings and the Newton constant, has
shown~\cite{ccm} that they do not exactly meet at a point; the error
being just a few percent. Hence, the big desert hypothesis is only
approximately valid, and one may expect new physics between
unification and present energy scales. Finally, NCSG predicts the
existence of a see-saw mechanism for neutrino masses with large
right-handed neutrino mass of the order of the cutoff scale $\Lambda$.

In conclusion, NCSG offers an elegant geometric interpretation of the
SM coupled to gravity. Applying Einstein's theory of General
Relativity (GR) within Riemannian geometry, one obtains the familiar
gravitational theory. As we have discussed above, applying the
spectral action approach within the context of an almost commutative
geometry, one gets gravity combined with Yang-Mills and Higgs. What
remains to be done in this programme, is to construct the appropriate
tools we need to apply within a fully noncommutative geometry and then
deduce the theory to which they will lead us.

\section{Physical meaning of the doubling of the algebra}
Let us highlight the physical implications of choosing an almost
commutative manifold. The geometry is
specified by the product ${\cal M}\times{\cal F}$ given from the
spectral triple
 \begin{equation}
({\cal A, H, D}, J, \gamma)=(C^\infty({\cal M}), L^2({\cal
     M},S),\slashed{\partial}_{\cal M}, J_{\cal
     M},\gamma_5)\otimes({\cal A_F, H_F, D_F}, J_{\cal F},\gamma_{\cal
     F})~, \nonumber
\end{equation}
defined as
\begin{equation}
({\cal A, H, D}, J, \gamma)=({\cal A}_1, {\cal H}_1, {\cal D}_1, J_1,
  \gamma_1)\otimes({\cal A}_2, {\cal H}_2, {\cal D}_2, J_2,
  \gamma_2)~, \nonumber\\
\end{equation}
with
\begin{equation}
{\cal A}={\cal A}_1\otimes{\cal A}_2~,~{\cal H}={\cal H}_1\otimes{\cal H}_2
~,~ {\cal D}={\cal D}_1\otimes 1 +\gamma_1\otimes{\cal D}_2~,
\gamma=\gamma_1\otimes\gamma_2~,~J=J_1\otimes J_2~,
\nonumber
\end{equation}
where $J^2=-1, [J,{\cal D}]=0, [J_1,\gamma_1]=0$ and $\{J,\gamma\}=0$.
The doubling of the algebra is intimately related to dissipation,
gauge field structure (necessary to address the physics of the SM), as
well as neutrino mixing, while it incorporates the seeds of
quantisation~\cite{PRD,mvpm}.

Consider the classical Brownian motion of a particle of mass $m$ with
equation of motion
\begin{equation}
m\ddot x(t)+\gamma\dot x(t)=f(t)~,
\end{equation}
where $f(t)$ denotes a random Gaussian distributed force. This
equation of motion can be derived from a Lagrangian in a canonical
procedure, using a delta functional classical constraint
representation as a functional integral.  It is easy to see~\cite{PRD}
that the constraint condition at the classical level introduces a new
coordinate, called $y$, with the $y$-system being the time-reversed of
the $x$-one, so that the equations of motion read
\begin{equation}
m\ddot x+\gamma\dot x=f~,~m\ddot y-\gamma\dot y=0~.
\end{equation}
The $x$-system represents an {\sl open} (dissipating) system, while
the $\{x,y\}$-system is a {\sl closed} one. This doubling, discussed
here in a completely classical context, is necessary in order to build
a canonical formalism for dissipative systems~\cite{PRD}.

To argue the relation between the doubling of the algebra and the
gauge field structure, consider
\begin{equation}
m\ddot x+\gamma\dot x+k x=0~,
\end{equation}
the equation of a classical one-dimensional damped harmonic
oscillator, with time independent quantities $m, \gamma, k$. Following
the previous discussion, we will complement the $x$-system with its
time-reversed image, called $y$-system, as
\begin{equation}
m\ddot y-\gamma\dot y+k y=0~,
\label{ampl}
\end{equation} 
in order to build a well-defined Lagrangian formalism. Equation
\eqref{ampl} above is that of a one-dimensional amplified harmonic
oscillator.

The Lagrangian of the closed $\{x,y\}$-system can be then written as
\begin{equation}
L={1\over 2m}(m\dot x_1+{e_1\over c}A_1)^2
-{1\over 2m}(m\dot x_2+{e_2\over c}A_2)^2
-{e^2\over 2mc^2}(A_1^2+A_2^2)-e\Phi~,
\end{equation}
where we have introduced the coordinates $x_1, x_2$ through
\begin{equation}
x_1(t)={x(t)+y(t)\over \sqrt 2}~,~ x_2(t)={x(t)-y(t)\over \sqrt 2}~,
\label{lagrangian}
\end{equation}
and the vector potential
\begin{equation}
A_i={B\over 2}\epsilon_{ij}x_j \ \ \mbox {for}\ i,j=1,2 ~\ \mbox
{with}\ B\equiv{\gamma c\over e}\ ,\ \epsilon_{ii}=0\ ,
\ \epsilon_{12}=-\epsilon_{21}=1~.
\end{equation} 
The Lagrangian \eqref{lagrangian} describes two particles having
opposite charges $e_1=-e_2\equiv e$ in the potential
$\Phi=\Phi_1-\Phi_2$, where $\Phi_i\equiv(k/2/e)x_i^2$ in the constant
magnetic field ${\bf B}={\bf\nabla}\times{\bf A}$.  Identifying the
doubled coordinate with the $x_2$, we observe that it acts as the
gauge field component $A_1$ to which the original $x_1$ coordinate is
coupled. We hence conclude that energy dissipated by one of the two
systems is gained by the other one, so that the gauge field can be
seen as the {\sl reservoir} in which the system is embedded.

Dissipation may also lead to a quantum evolution. This can be easily
shown by using 't Hooft's conjecture, saying that loss of information
(i.e., dissipation) in a regime of deterministic dynamics may lead to
a quantum mechanical evolution. We consider again the classical damped
harmonic $x$-oscillator and its time-reversed image, the
$y$-oscillator, discussed above. The Hamiltonian of the
$\{x,y\}$-system can be schematically written as
\begin{equation}
H=H_{\rm I}-H_{\rm II}~~\mbox{with the constraint} ~~
H_{\rm II}|\psi\rangle=0~,
\end{equation}
in order to define physical states $\psi$ and guarantee that the
Hamiltonian is bounded from below. The physical consequence of this
constraint is {\sl information loss}. Physical states are invariant
under time reversal and periodical, implying that
\begin{equation}
_H\langle \psi(\tau)|\psi(0)\rangle_H=e^{i\alpha \pi}~,
\end{equation}
where $\tau=2\pi/\Omega$ (with $\Omega$ expressed in terms of $m, k,
\gamma$) stands for the period, and $\alpha$ is a real constant.
Hence
\begin{equation}
\langle\psi_n(\tau)|H|\psi_n(\tau)\rangle=\hbar\Omega(n+\alpha/2)=\hbar\Omega
n+E_0~,
\end{equation}
with $E_0=(\hbar/2)\Omega\alpha$ the zero point ($n=0$) energy. Note
that the index $n$ above signals the $n$-dependence of the state and
the corresponding energy. In conclusion, the zero point quantum
contribution to the spectrum of physical states found above, results
from information loss, imposed by the underlying dissipative
dynamics~\cite{PRD}.

The algebra doubling can also lead to neutrino oscillations.  Linking
the algebra doubling to the deformed Hopf algebra, one can build
Bogogliubov operators as linear combinations of the co-product
operators defined in terms of the deformation parameter obtained from
the doubled algebra, and show the emergence of neutrino
mixing~\cite{mvpm}.  In particular, one can write the mixing
transformations connecting the flavour fields $\psi_{\rm f}$ to the
neutrino fields with nonvanishing masses $\psi_{\rm m}$ as
\begin{equation}
\nu_e(x)=G_\theta^{-1}(t)\nu_1(x)G_\theta(t)\ \ ;\ \ 
\nu_\mu(x)=G_\theta^{-1}(t)\nu_2(x)G_\theta(t)~,
\end{equation}
through the generator of field mixing transformations $G_\theta(t)$.
Note that for simplicity, and no loss of generality, we have only used
two neutrino species. Then writing $\psi_{\rm m}$ in terms of flavour
creation/annihilation operators, and similarly writing $\psi_{\rm m}$
in terms of mass creation/annihilation operators, one finds that
$G_\theta(t)$ contains rotation operator terms and Bogogliubov
transformation operator terms. Since deformed co-products are a basis
of Bogogliubov transformations, one concludes that field mixing arises
from the algebraic structure of the deformed co-product in the
noncommutative Hopf algebra embedded in the algebra doubling of
noncommutative spectral geometry. We can hence conclude that the SM
derived from NCSG, includes neutrino mixing by
construction~\cite{mvpm}.

\section{NCSG leading to an extended gravitational theory}

We are currently living in a very exciting time for early universe
cosmology, since our models can be now tested with a variety of very
precise astrophysical and high energy physics data, and in particular
with the Cosmic Microwave Background temperature anisotropies data and
the Large Hadron Collider results. However, despite the present golden
era of cosmology, a number of questions are still awaiting for a
definite answer. For instance, one does not know the origin of dark
matter and dark energy, whilst the search for a natural and
well-motivated inflationary model (or plausible alternatives to the
inflationary paradigm) still remains unsuccessful.

The main approaches to build early universe cosmological models have
been based to string/M-theory or some non perturbative approach to
Quantum Gravity, with Loop Quantum Cosmology being the leading
candidate.  Noncommutative spectral geometry can provide another
proposal, since the model lives by construction at the GUT scale.  

The bosonic action in Euclidean signature, favoured by the
formalism of spectral triples, is~\cite{ccm}
\beq\label{eq:action1} 
{\cal S}^{\rm E} = \int \left(
\frac{1}{2\kappa_0^2} R + \alpha_0
C_{\mu\nu\rho\sigma}C^{\mu\nu\rho\sigma} + \gamma_0 +\tau_0 R^\star
R^\star
\right.  
+ \frac{1}{4}G^i_{\mu\nu}G^{\mu\nu
  i}+\frac{1}{4}F^\alpha_{\mu\nu}F^{\mu\nu\alpha}\nonumber\\ 
+\frac{1}{4}B^{\mu\nu}B_{\mu\nu}
+\frac{1}{2}|D_\mu{\bf H}|^2-\mu_0^2|{\bf H}|^2
\left.
- \xi_0 R|{\bf H}|^2 +\lambda_0|{\bf H}|^4
\right) \sqrt{g} \ d^4 x~, \eeq
where 
\beq\label{bc}
\kappa_0^2=\frac{12\pi^2}{96f_2\Lambda^2-f_0\mathfrak{c}}~&,&~
\alpha_0=-\frac{3f_0}{10\pi^2}~,\nonumber\\
\gamma_0=\frac{1}{\pi^2}\left(48f_4\Lambda^4-f_2\Lambda^2\mathfrak{c}
+\frac{f_0}{4}\mathfrak{d}\right)~&,&~
\tau_0=\frac{11f_0}{60\pi^2}~,\nonumber\\
\mu_0^2=2\Lambda^2\frac{f_2}{f_0}-{\frac{\mathfrak{e}}{\mathfrak{a}}}~&,&~
\xi_0=\frac{1}{12}~,\nonumber\\
\lambda_0=\frac{\pi^2\mathfrak{b}}{2f_0\mathfrak{a}^2}~&,&~
{\bf H}=(\sqrt{af_0}/\pi)\phi~; \eeq
${\bf H}$ a rescaling of the Higgs field $\phi$ to normalize the
kinetic energy, and the momentum $f_0$ is physically related to the
coupling constants at unification. The geometric parameters
$\mathfrak{a,b,c,d,e}$ correspond to the (running) Yukawa parameters
of the particle physics model and the Majorana terms for the
right-handed neutrinos.  The first two terms in Eq.~(\ref{eq:action1})
depend only on the Riemann curvature tensor. The first is the
Einstein-Hilbert term and the second is the Weyl curvature
term; hence they are the Riemannian curvature terms.
The third one is the cosmological term, while the fourth term
\be R^\star
R^\star=\frac{1}{4}\epsilon^{\mu\nu\rho\sigma}\epsilon_{\alpha\beta\gamma\delta}
R^{\alpha\beta}_{\mu\nu}R^{\gamma\delta}_{\rho\sigma}~,\nonumber\ee
is the topological term that integrates to the Euler characteristic,
hence is nondynamical.  The three next terms are the Yang-Mills terms.
The eighth term is the scalar minimal coupling term, the next one is
the scalar mass term, and the last one is the scalar quartic potential
term. There is one more term, the $- \xi_0 R|{\bf H}|^2$, that couples
gravity with the SM. For $\xi_0=1/12$, this term encodes the conformal
coupling between the Higgs field and the Ricci curvature. 

Hence, the Lagrangian obtained through the NCSG approach contains, in
addition to the full SM Lagrangian, the Einstein-Hilbert action with a
cosmological term, a topological term related to the Euler
characteristic of the space-time manifold, a conformal Weyl term and a
conformal coupling of the Higgs field to gravity. Within the NCSG
context the Higgs field appears as the vector boson of the internal
noncommutative degrees of freedom.

At this point, let us make a few remarks. The relations given in
Eq.~(\ref{bc}) above, are tied to the cutoff scale $\Lambda$, hence
{\sl a priori} there is no reason for those to hold at any other
scale. Since the action Eq.~(\ref{eq:action1}) includes only the first
three terms in the asymptotic expansion, one must be cautious keeping
in mind that there are scales for which the neglected nonperturbative
effects become important. Since to study physical consequences of the
NCSG proposal one must use a Lorentzian signature,  we will
assume that a Wick rotation to imaginary time can be
achieved. Noticing the absence of quadratic terms in the
curvature --- there is only the term quadratic in the Weyl curvature and
the topological term $R^\star R^\star$ --- we immediately conclude that for
Friedmann-Lema\^{i}tre-Robertson-Walker geometries, the Weyl term
vanishes. Finally, notice the term that couples gravity with the SM, a
term which should always be present when one considers gravity coupled
to scalar fields.

The gravitational part of the asymptotic formula for the bosonic
sector of the NCSG, including the coupling between the Higgs field and
the Ricci curvature scalar, in Lorentzian signature, reads
\be\label{eq:1.5} {\cal S}_{\rm grav}^{\rm L} = \int \left(
\frac{1}{2\kappa_0^2} R + \alpha_0
C_{\mu\nu\rho\sigma}C^{\mu\nu\rho\sigma} + \tau_0 R^\star
R^\star\right.  
 -\left.  \xi_0 R|{\bf H}|^2 \right)
\sqrt{-g} \ d^4 x~.\ee
It will lead to the following equations of motion~\cite{Nelson:2008uy}:
\be\label{eq:EoM2} R^{\mu\nu} - \frac{1}{2}g^{\mu\nu} R +
\frac{1}{\beta^2} \delta_{\rm cc}\left[
  2C^{\mu\lambda\nu\kappa}_{;\lambda ; \kappa} +
  C^{\mu\lambda\nu\kappa}R_{\lambda \kappa}\right]= 
\kappa_0^2 \delta_{\rm cc}T^{\mu\nu}_{\rm matter}~, \ee
where
\be
\beta^2 \equiv -\frac{1}{4\kappa_0^2 \alpha_0}
\ \ \ \ \mbox{and}\ \ \ \
\delta_{\rm cc}\equiv[1-2\kappa_0^2\xi_0{\bf H}^2]^{-1}~.
\ee
The definition $\delta_{\rm cc}$ captures the conformal coupling
between the Ricci scalar and the Higgs field.

In the low energy weak curvature regime, one may neglect the
nonminimal coupling term between the background geometry and the Higgs
field, getting $\delta_{\rm cc}=1$.  Hence, for a cosmological
context, namely Friedmann-Lema\^{i}tre-Robertson-Walker space-time,
the Weyl tensor vanishes and the noncommutative spectral geometry
corrections to the Einstein equation vanish~\cite{Nelson:2008uy}.
Consequently, any modifications to the background equation may appear
at leading order only for anisotropic and inhomogeneous models, such as
a Bianchi type-V model defined by the space-time metric
\be
g_{\mu\nu} = {\rm diag} \left[ -1,\{a_1(t)\}^2e^{-2nz} ,
  \{a_2(t)\}^2e^{-2nz}, \{a_3(t)\}^2 \right]~, \ee
where $a_i(t)$ with $i=1,2,3$, arbitrary functions, denoting the three
scale factors, and $n$ an integer.  In this metric, the modified
Friedmann equation reads~\cite{Nelson:2008uy}:
\beq\label{eq:Friedmann_BV} \kappa_0^2 T_{00}=&&\nonumber\\
 - \dot{A}_3\left(
\dot{A}_1+\dot{A}_2\right) -n^2 e^{-2A_3} \left( \dot{A}_1
\dot{A}_2-3\right)
&& \nonumber \\
 +\frac{8\alpha_0\kappa_0^2 n^2}{3} e^{-2A_3} \left[
  5\left(\dot{A}_1\right)^2 + 5\left(\dot{A}_2\right)^2 -
  \left(\dot{A}_3\right)^2\right.
&&\nonumber\\
\left. - \dot{A}_1\dot{A}_2 - \dot{A}_2\dot{A}_3
  -\dot{A}_3\dot{A}_1 - \ddot{A}_1 - \ddot{A}_2 - \ddot{A}_3 + 3
  \right]
- \frac{4\alpha_0\kappa_0^2}{3} \sum_i \Biggl\{
\dot{A}_1\dot{A}_2\dot{A}_3 \dot{A}_i
&&\nonumber\\
 + \dot{A}_i \dot{A}_{i+1} \left(
\left( \dot{A}_i - \dot{A}_{i+1}\right)^2 -
\dot{A}_i\dot{A}_{i+1}\right)
 + \left( \ddot{A}_i + \left( \dot{A}_i\right)^2\right)\left[
  -\ddot{A}_i - \left( \dot{A}_i\right)^2 + \frac{1}{2}\left(
  \ddot{A}_{i+1} + \ddot{A}_{i+2} \right)\right.&&\nonumber\\
\left. + \frac{1}{2}\left(
  \left(\dot{A}_{i+1}\right)^2 + \left( \dot{A}_{i+2}\right)^2 \right)
  \right]
%
+ \left[ \dddot{A}_i + 3 \dot{A}_i \ddot{A}_i -\left( \ddot{A}_i +
  \left( \dot{A}_i\right)^2 \right)\left( \dot{A}_i - \dot{A}_{i+1} -
  \dot{A}_{i+2} \right)\right]
&&\nonumber\\
\times\left[ 2\dot{A}_i
  -\dot{A}_{i+1}-\dot{A}_{i+2} \right]\Biggr\}~, \eeq
where we have defined
$A_i\left(t\right) = {\rm ln} a_i\left(t\right)$  with $i=1,2,3$.
One then immediately observes that the correction terms in
Eq.~(\ref{eq:Friedmann_BV}) above come in two types. Those which are
fourth order in time derivatives, and those that are at the same order
as the ones derived from the standard Einstein-Hilbert action. The
former can be considered as small corrections since usually in
cosmology we have slowly varying functions.  The latter are
proportional to $n^2$, hence they vanish for homogeneous versions of
Bianchi type-V. In conclusion, the corrections to Einstein's equations
are only present in inhomogeneous and anisotropic
space-times~\cite{Nelson:2008uy}.

As energies are approaching the Higgs scale, the nonminimal coupling
of the Higgs field to the curvature can no longer be neglected.  Then
the equations of motion read~\cite{Nelson:2008uy}
\be R^{\mu\nu} - \frac{1}{2}g^{\mu\nu}R =
\kappa_0^2\left[\frac{1}{1-\kappa_0^2 |{\bf H}|^2/6}\right] T^{\mu\nu}_{\rm
  matter}~, \ee 
where for simplicity we set $\beta=0$.  We thus conclude that $|{\bf
  H}|$ leads to an effective gravitational constant.  A different way
to see the r\^ole of the nonminimal coupling is by examining its
effect on the equations of motion for the Higgs field in
a constant gravitational field.  Hence, since
\be -\mu_0 |{\bf H}|^2 \rightarrow -\left( \mu_0 + \frac{R}{12}\right)
|{\bf H}|^2~,  \ee
we conclude that for constant curvature, the self interaction of the
Higgs field is increased.

Finally, one may find links to dilatonic gravity and chameleon
cosmology~\cite{Nelson:2008uy}. Firstly, the action 
\be\label{eq:action_Higgs} {\cal L}_{|{\bf H}|} = -\frac{R}{12}|{\bf
  H}|^2 + \frac{1}{2} |D^\alpha {\bf H} | | D ^\beta {\bf H} |
g_{\alpha\beta} - \mu_0 |{\bf H}|^2 + \lambda_0|{\bf H}|^4 \ee
(where $D^\alpha$ denotes covariant derivative) for the pure Higgs
field ${\bf H}$, can be written in the form of four-dimensional
dilatonic gravity as
\be {\cal L}_{\tilde{\phi}} = e^{-2\tilde{\phi}} \left[ -R + 6D^\alpha
  \tilde{\phi}D^\beta \tilde{\phi} g_{\alpha\beta} - 12\left( \mu_0
  -12\lambda_0 e^{-2\tilde{\phi}}\right) \right]~, \ee
by a redefinition of the Higgs field as
\be \tilde{\phi} = -\ln \left( |{\bf H}|/(2\sqrt{3})\right)~.
\ee
Secondly, chameleon models are characterised by the existence of a
scalar field having a nonminimal coupling to the standard matter
content (thus evading solar system tests of GR).  In the context of
NCSG, the Higgs field has a nonzero coupling to the background
geometry. If the equations of motion can be approximated by Einstein's
equations, then the background geometry will be approximately given by
standard matter, making the mass and dynamics of the Higgs field
explicitly dependent of the local matter content.

Exploring the possible r\^ole of scalar fields appearing in the NCSG
action, one may wonder whether the Higgs field, through its nonminimal
coupling to the background geometry, can be the inflaton. The
Gravity-Higgs sector of the asymptotic expansion of the spectral
action, in Lorentzian signature reads
\be
S^{\rm
  L}_{\rm GH}=\int\Big[\frac{1-2\kappa_0^2\xi_0
    H^2}{2\kappa_0^2}R 
-\frac{1}{2}(\nabla  H)^2- V(H)\Big] \sqrt{-g}\  d^4x~,
\ee
where 
\be\label{higgs-pot}
V(H)=\lambda_0H^4-\mu_0^2H^2~,
\ee
with $\mu_0$ and $\lambda_0$ subject to radiative corrections as
functions of energy.  For each value of the top quark mass, there is a
value of the Higgs mass where the Higgs potential is locally
flattened~\cite{mmm}.  However, since the flat region is narrow, the
slow-roll must be very slow, otherwise the quasi-exponential expansion
will not last long enough.  Moreover, the amplitude of density
perturbations in the Cosmic Microwave
Background must be in agreement with the measured one. 

Calculating the renormalisation of the Higgs self-coupling up to
two-loops and constructing an effective potential which fits the
renormalisation group improved potential around the flat region, one
concludes that while the Higgs potential can lead to the slow-roll
conditions being satisfied, the constraints imposed from the CMB data make
the predictions of such a scenario incompatible with the measured
value of the top quark mass~\cite{mmm}.

The gravitational sector of the NCSG action provides a proposal for an
extended theory of gravity. Studying the astrophysical consequences of
such a theory, one is able to constrain one of the three momenta of
the cutoff function, namely $f_0$ (or equivalently $a_0=
-3f_0/(10\pi^2)$ in Eq.~(\ref{eq:1.5})) which specifies the initial
conditions on the gauge couplings~\cite{Nelson:2010ru}-\cite{sgaam}.
Hence we will get a restriction on the particle physics at
unification.  Note that one cannot constrain the other two free
parameters (the momenta $f_2, f_4$) without an {\sl ad hoc} assumption
on the running of the coefficients in the action functional.

To simplify the analysis and with no loss of generality, let us
neglect in the following the conformal coupling between the Ricci
curvature and the Higgs field. Hence to find nonzero correction terms
we have to go beyond the homogeneous and isotropic case. The equations
of motion read~\cite{Nelson:2008uy} 
\be
G^{\mu\nu}+\frac{1}{\beta^2}[2\nabla_\lambda \nabla_\kappa
  C^{\mu\nu\lambda\kappa}+C^{\mu\lambda\nu\kappa}R_{\lambda\kappa}]
=\kappa^2 T^{\mu\nu}_{({\rm matter})}\,, 
\ee
where $\kappa^2\equiv 8\pi G$, $G^{\mu\nu}$ is the (zero order)
Einstein tensor, $T^{\mu\nu}_{\rm matter}$ the energy-momentum tensor
of matter and $\beta^{2} =\displaystyle{5\pi^2/(6\kappa^2f_0)}$. 
Performing a detailed analysis of linear perturbations
\be
g_{\mu\nu}=\eta_{\mu\nu}+\gamma_{\mu\nu}~, 
\ee
around a Minkowski background metric $\eta_{\mu\nu}$, one can show
that the linearised equation of motion, derived within the NCSG
context, reads~\cite{Nelson:2008uy}
\begin{equation}\label{waveeq}
    \left(1-\frac{1}{\beta^2}\Box_\eta \right)\Box_\eta {\bar
    h}^{\mu\nu}= - 2\kappa^2 T^{\mu\nu}_{\rm matter}\,,
\end{equation}
with $T^{\mu\nu}_{\rm matter}$ taken to lowest order in
$\gamma^{\mu\nu}$, so that it is independent of $\gamma^{\mu\nu}$ and
satisfies the conservation equation $\partial_\mu T^{\mu\nu}_{({\rm
    matter})}=0$. The equation of motion above has been written in
terms of the tensor~\cite{Nelson:2010rt}
\begin{equation}\label{bar-h}
    {\bar h}_{\mu\nu}={\bar \gamma}_{\mu\nu}-\frac{1}{3\beta^2}\,
    {\cal
      Q}^{-1}\left(\eta_{\mu\nu}\Box_\eta-\partial_\mu\partial_\nu\right)\gamma\,,
\end{equation}
with
\be
\label{def-Q}
{\cal Q}\equiv 1-\frac{1}{\beta^2}\, \Box_\eta~, 
\ee 
and having defined ${\bar \gamma}_{\mu\nu}$ the {\sl trace reverse} of
$\gamma_{\mu\nu}$:
\be
{\bar \gamma}_{\mu\nu}=\gamma_{\mu\nu}-{1\over 2}\eta_{\mu\nu}\gamma~.
\ee
Note that we constrain $\alpha_0<0$, hence $\beta^2>0$, so that
Minkowski is a stable vacuum of the theory. The general
solution~\cite{Nelson:2010rt}
\beq \label{eq:field} h^{\mu\nu} = 2\beta^2\kappa\int {\rm d}S(x')
G_{\rm R}(x,x') T^{\mu\nu}(x')\, , \eeq
to Eq.~(\ref{waveeq}) is given in terms of Green's
functions $G_{\rm R}(x,x')$ which satisfy the 
fourth-order partial differential equation
\beq
\label{gf:1}
\Big (\Box-\beta^2\Big)\Box G_{\rm R}(x,x') \, &=& 4\pi\delta^{(4)}(x-x') \, ;
\eeq
the operators $\Box$ above are acting on $x$. Consequently, the field
is given by~\cite{Nelson:2010rt}
\be h^{\mu\nu}\left({\bf r},t\right) =\frac{4G\beta}{c^4} \int {\rm
  d}{\bf r}' {\rm d} t'\frac{\Theta\left(T\right)}{\sqrt{\left(cT\right)^2
    - |{\bf R}|^2}}{\cal J}_1\left(
\beta\sqrt{\left(cT\right)^2 - |{\bf R}|^2}\right) T^{\mu\nu}\left(
{\bf r}',t'\right)\Theta\left( cT - |{\bf R}|\right) ~,
\ee
where $J_1(x)$ is the first order Bessel function of the first kind;
$\Theta$ is the Heavyside step function; $T=t-t'$ is the difference
between the time of observation $t$ and time of emission $t'$ of the
perturbation; and ${\bf R} ={\bf r}-{\bf r}'$ denotes the difference
between the location ${\bf r}$ of the observer and the location ${\bf
  r}'$ of the emitter.  One can then calculate the propagation of
gravitational waves and investigate discrepancies from the results
obtained within standard GR.

In the far-field limit $|{\bf r}| \approx |{\bf r} - {\bf r}'|$, the spatial
components of the $h^{\mu\nu}$ field are~\cite{Nelson:2010rt}
\be\label{eq:4} 
h^{ik}\left({\bf r},t\right) \approx \frac{2G
  \beta}{3c^4} \int_{-\infty}^{t-\frac{1}{c}|{\bf r}|} \frac{{\rm
    d}t'}{\sqrt{c^2\left( t-t'\right)^2 - |{\bf r}|^2} }
{\cal J}_1 \left(
\beta\sqrt{c^2\left( t-t'\right)^2 - |{\bf r}|^2}\right)
\ddot{D}^{ik}\left(t'\right)~, 
\ee
where $D^{ik}$ stands for the quadrupole moment,
\be D^{ik}\left(t\right) \equiv \frac{3}{c^2}\int {\rm d}{\bf r} \ 
x^i x^k T^{00}({\bf r},t)~.
\ee
In this limit, the rate of energy loss from a circular binary system,
under the assumption that the internal structure of the pair of masses
$m_1$ and $m_2$ can be neglected, reads
\be\label{eq:energy} -\frac{{\rm d} {\cal E}}{{\rm d}t} \approx
\frac{c^2}{20G} |{\bf r}|^2 \dot{h}_{ij} \dot{h}^{ij}~,  \ee
where the time derivatives of the spatial components of the field are
\be\label{eq:grav_pert} 
\dot{h}^{ij} = \frac{4G\beta
  A^{ij}\omega^{ij}}{3c^4} \Biggl[ \sin\left( \omega^{ij} t +
  \phi^{ij} \right) f_c\left( \beta |{\bf r}|,
  \frac{\omega^{ij}}{\beta c}\right)+ \cos\left(
  \omega^{ij}t+\phi^{ij}\right) f_s\left( \beta |{\bf r}|,
  \frac{\omega^{ij}}{\beta c}\right) \Biggr]~, \ee
(note that no summation is implied) and we have defined the functions
$ f_{\rm s}\left( x,z\right), f_{\rm c}\left( x,z\right)$ as
\beq\label{eq:f1}
 f_{\rm s}\left( x,z\right) &\equiv& \int_0^\infty
\frac{{\rm d}s}{\sqrt{s^2 + x^2}} {\cal J}_1\left(s\right) \sin
\left(z\sqrt{ s^2 + x^2} \right)~,\nonumber\\
\label{eq:f2}
f_{\rm c}\left( x,z\right) &\equiv&
\int_0^\infty \frac{{\rm d}s}{\sqrt{s^2 + x^2}} {\cal
  J}_1\left(s\right) \cos \left(z\sqrt{ s^2 + x^2} \right)~.
\eeq
The integrals in Eq.~(\ref{eq:f2}) show a strong resonance behaviour
at $z=1$ corresponding to a critical frequency
\be
2\omega_{\rm c}=c\beta =c(-\alpha_0 G)^{-1}~,
\ee
close to which strong deviations from the standard results of GR are
expected; these integrals are easily evaluated numerically for
$z>1$ and $z<1$.

In the large $|{\bf r}|$ limit, the rate of energy loss to
gravitational radiation by a circular (for simplicity) binary system
of masses $m_1, m_2$ (we denote by $\mu$ the reduced mass of the
system) at a separation vector of magnitude $\rho$
reads~\cite{Nelson:2010ru} \beq\label{eq:GR_approx} -\frac{{\rm d}
  {\cal E}}{{\rm d}t} \approx \frac{32 G \mu^2 \rho^4 \omega^6}{5c^5}
\times\left\{ \begin{array}{cc} 1 + \frac{C}{\beta|{\bf r}| \left( 1-
    \frac{\omega}{\omega_{\rm c}}\right)} {\cal J}_1 \left( \beta
  |{\bf r}| - \frac{\omega}{\omega_{\rm c}}\right) +\dots & ; \omega <
  \omega_{\rm c} \\ 4\sin^2\left( \beta|{\bf r}|
  \tilde{f}\left(\frac{\omega}{\omega_{\rm c}}\right)\right) & ;
  \omega > \omega_{\rm c}
\end{array} \right.~,
\eeq
where in the $\omega < \omega_{\rm c}$ case the dots refer to higher
powers of $1/\left(\beta |{\bf r}|\right)$.  

Hence, for orbital frequencies small compared to the critical one
$\omega_{\rm c}$, any deviation from the standard GR result is
suppressed by the distance to the source. In this case, the
$\beta\rightarrow \infty$ limit leads, as expected, to the GR
result. This is not the case however for $\omega>\omega_{\rm c}$,
since then the GR result in only recovered if $\beta |{\bf
  r}|\tilde{f}\left(\omega/\omega_{\rm c}\right) = \pi/3$.  We will
thus only consider the $\omega < \omega_{\rm c}$ physically
interesting case, and restrict $\beta$ (equivalently $f_0$) by
requiring that the energy lost to gravitational radiation agrees with
the one predicted by GR to within observational uncertainties.  Note that
the presence of the Bessel function implies that the amplitude of the
deviation from the result obtained within standard GR will oscillate
with frequencies as well as distances, however the
effect will be suppressed by the $|{\bf r}|^{-1}$ factor.

Considering binary pulsar systems, for which the rate of change of the
orbital frequency is well known, the observational constraint is
$\beta \gtrsim 7.55\times 10^{-13} {\rm m}^{-1}$~\cite{Nelson:2010ru}.
This (weak) limit can be improved through future observations of
rapidly orbiting binaries relatively close to the Earth.

One can also set a lower bound on the Weyl term appearing in the NCSG
action using results from the Gravity Probe B~\cite{gam} and LAser
RElativity Satellite (LARES)~\cite{sgaam} experiments.  Gravity Probe
B satellite contains a set of four gyroscopes and has tested the
geodesic and the frame-dragging (Lense-Thirring) effects of GR with
extreme precision. The LARES mission is designed to test these two
effects to within $1\%$ of the value predicted within the theory of
GR.

Let us write the metric in terms of the metric potentials $\Phi,
\Psi$ and the vector potential ${\bf A}$ as
 \begin{equation}\label{elementline}
    ds^2 = -(1+2\Phi)dt^2+ 2{\bf A}\cdot d{\bf x} dt+(1+2\Psi) d{\bf x}^2 \,.
 \end{equation}
 The rate of an orbiting gyroscope precession can be then splitted
 into a part generated by the metric potentials and one generated by
 the vector potential. The obtained spin equation of motion for the
 gyro-spin three-vector is hence expressed as the sum of the
 instantaneous geodesic and Lense-Thirring precessions. Each of these
 two precessions can be then written as the sum of two terms, one obtained
 within GR and one coming from NCSG. Setting the geodesic precession
 (equivalently for the Lense-Thirring precession) to be the one
 predicted from standard GR and requiring that the NCSG contribution
 is within the accuracy of its measured value, Gravity Probe B
 results imply~\cite{gam} $\beta\gtrsim 7.1\times 10^{-5} {\rm
   m}^{-1}$, and LARES experiment sets~\cite{sgaam}
 $\beta\gtrsim 1.2\times 10^{-6} {\rm m}^{-1}$.

A much stronger constraint can be imposed to $\beta$ using the torsion
balance experiments. The modifications induced by the
NCSG action to the Newtonian potentials $\Phi, \Phi$ lead to the
following expressions for the components of $\gamma_{\mu\nu}$~\cite{gam}:
\begin{eqnarray}\label{gamma001}
    \gamma_{00}&=&-2\Phi=\frac{2GM}{r}\left(1-\frac{4}{3}e^{-\beta
    r}\right)\,, \nonumber \\ \gamma_{0i}&=&\gamma_{i0}= A_i 
=-\frac{4G}{r^3}[1-(1+\beta r)e^{-\beta r}]({\bf r}\wedge
    {\bf J})_i\,, \nonumber \\ \gamma_{ij} &=& 2\Psi \delta_{ij}
    = \frac{2GM}{r}\left[1+\frac{5}{9}e^{-\beta
    r}\right]\delta_{ij}\,.
 \end{eqnarray}
These modifications are similar to those induced by a
fifth-force through a potential
\be\label{fifthV}
V(r)=-\displaystyle{\frac{GMm}{r}\Big(1+\alpha e^{-r/\lambda}\Big)}~,
\ee
where $\alpha$ is a dimensionless strength parameter and $\lambda$ a
length scale.  The tightest constraint on $\lambda$ from the latest
torsion balance experiments is $\lambda\lesssim 10^{-4}\mbox{m}$
leading to~\cite{gam}
\begin{equation}\label{constr}
 \beta \gtrsim 10^4 \mbox{m}^{-1}\,,
 \end{equation}
a much stronger constraint than the ones obtained through pulsar
timings, Gravity Probe B or LARES experiments.

\section{A new approach: The zeta function regularisation}
The cutoff bosonic spectral action is a quite successful and
promising scheme, worth to be further investigated. Based upon an
elegant mathematical theory, it offers a description of geometry in
terms of spectral properties of operators and leads to a model of
particle interactions which is very close to the real phenomenology as
revealed from high energy physics experiments. Whilst the Standard
Model and the Pati-Salam gauge groups fit into the NCSG model, the
SU(5) or SO(10) groups do not; and absence of large groups is
interesting since it prevents proton decay. Following the NCSG scheme,
one is able to infer quantities related to the Higgs boson based only
upon the input from the fermionic parameters in the fluctuated Dirac
operator ${\cal D}_A$, which defines also the fermionic part of the
bosonic spectral action. 

However, despite its success, the cutoff spectral action faces some
issues. It is calculated via the asymptotic heat kernel expansion and
is only valid when the fields and their derivatives are small with
respect to the cutoff energy scale $\Lambda$. Hence, the asymptotic
expansion leading to the appearance of only three of the momenta of
the cutoff function $f$ is only valid in the weak-field approximation
and one may wonder what does it happen in the ultraviolet regime when
high momenta are the dominant ones. Issues with
super-renormalisability have been addressed in the literature,
indicating that high energy bosons do not propagate~\cite{klv}. In
short, within the traditional cutoff bosonic spectral action approach,
it is not clear what is the meaning of the cutoff scale $\Lambda$ nor
what does it happen at scales beyond $\Lambda$. Moreover, the cutoff
bosonic spectral action depends (even though not in a very strong
matter) on the particular choice of the cutoff $f$ function. Finally,
there is an issue with the magnitude of the dimensionful parameters in
the model, namely the cosmological constant, the Higgs vacuum
expectation value and the gravitational coupling. The natural value
for the cosmological constant obtained through the spectral action
approach is $\sim\Lambda^4$, which is clearly much bigger than its
observational value. Thus, to render it compatible with the
observational value of the cosmological constant, one should add by hand
an appropriate term. The heat expansion does not lead to a minimum of
the Higgs potential for all natural choices of the cutoff
function. Hence one must add by hand to the $H^2$ term, a quadratic
term with a large coefficient, to provide a minimum of the potential
with a Higgs vacuum expectation value which is many orders of
magnitude smaller than $\Lambda$. There is also a problem with the
value of the gravitational constant given by the coefficient in front
of the scalar curvature. The value obtained through the cutoff bosonic
spectral action is at least one order of magnitude smaller than its
experimental value. Therefore, one has again to add by hand an
appropriate term. In conclusion, the physical values of these
dimensionful parameters necessitate an experimental input beyond the
NCSG approach. A different way of phrasing this problem is by calling
it {\sl the naturalness problem}.

To cure the dependence on the cutoff scale and the cutoff function, it
has been recently proposed~\cite{mfma} another way to regularise the
infinite sum of the eigenvalues of the (unbounded) fluctuated Dirac
operator, based on the $\zeta$ function. More precisely, the zeta
bosonic spectral action can be defined as~\cite{mfma} 
\be S_\zeta \equiv \lim_{s\rightarrow 0} \Tr {\cal D}^{-2s}\equiv
\zeta(0,{\cal D}^2)~, \label{szetadef} \ee
with the zeta function given by the $a_4$ heat kernel coefficient
associated with the Laplace type operator ${\cal D}^2$:
\be S_\zeta = a_4\left[{\cal D}^2\right] = \int d^4 x \,\sqrt{g}\, L ~~
\mbox{with} ~~ L(x) = a_4({\cal D}^2,x)~. \label{BSAv20} \ee
This $\zeta$ spectral function leads to the following Lagrangian
density~\cite{mfma}:
\beq L(x) &=& \alpha_1M^4 + \alpha_2 M^2 R +\alpha_3M^2H^2
+ \alpha_4 B_{\mu\nu}B^{\mu\nu} + \alpha_5
W_{\mu\nu}^{\alpha}W^{\mu\nu\,\alpha} + \alpha_6
G_{\mu\nu}^aG^{\mu\nu\,a} \nonumber\\ && +
\alpha_7\,H\left(-\nabla^2-\frac{R}{6}\right)H + \alpha_8 H^4 +
\alpha_9 C_{\mu\nu\rho\sigma}C^{\mu\nu\rho\sigma} +
\alpha_{10}R^*R^*~. \label{L} \eeq
Note that $B_{\mu\nu}$, $W_{\mu\nu}$ and $G_{\mu\nu}$ stand for the
field strength tensors of the corresponding U(1), SU(2) and SU(3)
gauge fields; $\alpha_{1},..,\alpha_{10}$ are dimensionless constants
determined by the Dirac operator; $R^*R^*$ is the Gauss-Bonnet density
and $C$ is the Weyl tensor.

Up to this point, the dimensionful quantity $M$ appearing in the
position corresponding to the Majorana mass of the right-handed
neutrino in the Dirac operator, is just a constant. All dimensionless
constants are normalised to their spectral action values, whilst the
three dimensionful parameters are normalised from experiments. In
analogy with the cutoff bosonic spectral action we consider the zeta
bosonic spectral action as valid at the scale $\Lambda\sim
(10^{14}-10^{17})\ {\rm GeV}$, but emphasising that the action is itself
independent of the scale $\Lambda$.
Since the zeta spectral action Eq.~(\ref{L}) does not contain higher
than 4-dimensional operators, it is renormalisable and
local. Moreover, there are no issues about asymptotic expansion or
convergence, and the zeta spectral action is purely spectral with
no dependence on a cutoff function. 

Within the cutoff spectral action approach, if the full
momentum-dependence of the propagators is considered, then the
spectral dimension becomes independent of the spin and vanishes
identically~\cite{Alkofer:2014raa}. A physical interpretation of this
behaviour can be summarised in the statement that {\sl high energy
  bosons do not propagate}~\cite{klv}.  However, the zeta spectral
action leads to nontrivial spectral dimensions.  The Higgs scalar part
of the bosonic action has the same bahaviour in the ultraviolet as in
the infrared limit, so that the spectral dimension coincides with the
topological dimension of the manifold. The same holds for the gauge
fields. Hence the spectral dimension of the Higgs scalar and the gauge
fields is equal to four. To calculate the gravitational spectral
dimension one needs to do an analytic continuation~\cite{mfma}; we
will highlight the computation below.

The spectral dimension $D_{\rm s}$ is defined as
\be D_{\rm s} \equiv \lim_{T\rightarrow 0}\left[-2 \frac{\partial \log
    P(T)}{\partial \log T}\right]~, \label{spdim} \ee
where $P(T)$ stands for the value of the heat kernel $P(T,x,x')$,
corresponding to the quadratic part of the gravitational part of the
action for transverse and traceless fluctuations $h_{\mu\nu}$ of the
metric tensor $g_{\mu\nu}$, at $x=x'$, and $T$ denotes the diffusion
time.  Such a heat kernel is given by the integral
\be P(T,x,x') = \int \frac{d^4 p}{(2\pi^4)}e^{ip(x-x')}
e^{-\left(p^2 - a p^4\right)T}~,\label{hk} \ee
and one can show that there exists an analytic continuation of the
relevant integral in a region of positive $a$~\cite{mfma}. The
obtained gravitational spectral dimension for all nonzero real $a$
is~\cite{mfma}
\be
D_{\rm s} = 2~; \label{res}
\ee
in agreement with the fact that here the gravitational propagators
decrease faster in infinity due to the presence of fourth derivatives.
Finally, there exists a {\sl low energy} limit of the gravitational
spectral dimension, valid for all real $a$, with~\cite{mfma}
\be
D_{\rm s}^{\rm low} = 4~,\label{lowres}
\ee
as expected since at very low energies the dynamics does not feel the
fourth derivative terms.

\section{Conclusions}
Noncommutative spectral geometry offers a geometric framework for the
description of the Standard Model of strong and electroweak
interactions, based upon a purely algebraic description. The
constructed spectral action for fermions and bosons based on the
spectral properties of the generalised Dirac operator and for an
appropriately chosen algebra led to phenomenological results very
close to the experimental ones.

The doubling of the algebra, which can be interpreted as considering a
geometric space formed by two copies of a four-dimensional manifold,
has profound physical implications.  In particular, the doubling of
the algebra is required in order to incorporate gauge symmetries, a
fundamental ingredient of the Standard Model, and it is also the main
element to explain neutrino mixing. Following 't Hooft's conjecture,
one can also show that the NCSG classical construction carries
implicit in its doubling of the algebra the seeds of quantisation.

Considering the gravitational sector of the spectral action, one can
constrain one of the momenta of the cutoff function, namely the one
related to the coupling constants at unification. The strongest
constraint on the coefficient of the curvature squared term is obtained
through torsion balance experiments, while using data from binary
pulsars, GPB and LARES experiments the corresponding constraint is
much weaker.

To address the issues of renormalisability and spectral dimensions, a
novel definition of the bosonic spectral action has been proposed,
based upon the $\zeta$ function regularisation. While the zeta
spectral action shares the same predictive power with the traditional
cutoff spectral action, only the former leads to a local, unitary and
renormalisable theory. In this new proposal, the open aspect that
remains to be addressed is the dynamical generation of the three
dimensionful fundamental constants, namely the cosmological constant,
the Higgs vacuum expectation value, and the gravitational constant.


\medskip
\begin{thebibliography}{14}
\bibitem{ncg-book1} Connes A 1994 {\it Noncommutative Geometry} (New
  York: Academic Press)

\bibitem{ncg-book2} Connes A and Marcolli M 2008 {\it Noncommutative
  Geometry, Quantum Fields and Motives} (India: Hindustan Book Agency)

\bibitem{ccm} Chamseddine A H, Connes A and Marcolli M 2007 {\it
  Adv. Theor. Math. Phys.} {\bf 11} 991

\bibitem{Walterbook} van Suijlekom W D 2014 {\it Mathematical Physics
  Studies Springer}; ISBN: 978-94-017-9161-8 (Print) 978-94-017-9162-5
  (Online)
 a geometric space fromed by
two copies of a four-dimensional manifold
\bibitem{Sakellariadou:2011dk} Sakellariadou M 2010 {\it PoS CNCFG}
  {\bf 2010} 028

\bibitem{Sakellariadou:2010nr} Sakellariadou M 2011 {\it
  Int.\ J.\ Mod.\ Phys.} D {\bf 20} 785

\bibitem{Sakellariadou:2012jz} Sakellariadou M 2011 {\it PoS CORFU}
  {\bf 2011} 053

\bibitem{PRD} Sakellariadou M, Stabile A and Vitiello G 2011 {\it
  Phys. Rev.} D {\bf 84} 045026

\bibitem{mvpm} Gargiulo M V, Sakellariadou M and Vitiello G
2014 {\it Eur.\ Phys.\ J.} C {\bf 74} 2695

\bibitem{Nelson:2008uy} Nelson W and Sakellariadou M 2010 {\it
Phys.\ Rev.} D {\bf 81} 085038

\bibitem{Nelson:2009wr} Nelson W and Sakellariadou M 2009 {\it
Phys.\ Lett.} B {\bf 680} 263

\bibitem{Marcolli:2009in} Marcolli M and Pierpaoli E 2010 {\it
  Adv.\ Theor.\ Math.\ Phys.} {\bf 14}

\bibitem{mmm} Buck M, Fairbairn M and Sakellariadou M 2010 {\it
Phys.\ Rev.} D {\bf 82} 043509

\bibitem{Nelson:2010ru} Nelson W, Ochoa J and Sakellariadou M 2010
{\it Phys.\ Rev.\ Lett.} {\bf 105} 101602

\bibitem{Nelson:2010rt} Nelson W, Ochoa J and Sakellariadou M 2010
{\it Phys.\ Rev.} D {\bf 82} 085021

\bibitem{Sakellariadou:2010nr} Sakellariadou M 2011 {\it
Int. J. Mod. Phys.} D {\bf 20} 785

\bibitem{gam} Lambiase G, Sakellariadou M and Stabile A 2013
{\it J.\ Cosm.\ \& Astrop.\ Phys.} {\bf 12} 020

\bibitem{sgaam} Capozziello S, Lambiase G, Sakellariadou M, Stabile A
  and Stabile An 2015 {\it Phys.\ Rev.} D {\bf 91} 044012

\bibitem{mfma}
Kurkov M A, Lizzi F, Sakellariadou M and Watcharangkool A 2015
{\it Phys.\ Rev.} D {\it in press}
 {\it Preprint} arXiv:1412.4669 [hep-th].

\bibitem{Chamseddine:2007ia} Chamseddine A H and Connes A 2007 {\it
Phys. Rev. Lett.} {\bf 99} 191601

\bibitem{ac1997} Chamseddine A H and Connes A 1997
 {\it Comm.\ Math.\ Phys.} {\bf 186} 731 

\bibitem{Resilience}
Chamseddine A H and Connes A 2012
{\it  JHEP} {\bf 1209} 104

\bibitem{Stephan} 
Stephan C A 2009
{\it Phys.\ Rev.} D {\bf 79} 065013

\bibitem{grand-algebra}
Devastato A, Lizz Fi and Martinetti P 2014
 {\it JHEP} {\bf 1401} 042

\bibitem{Chamseddine:2013sia}
Chamseddine A H, Connes A and van Suijlekom W D 2013
{\it J.\ Geom.\ Phys.}  {\bf 73} 222

\bibitem{klv} Kurkov M A, Lizzi F and Vassilevich D 2014 {\it
  Phys.\ Lett.} B {\bf 731} 311

\bibitem{Alkofer:2014raa}
Alkofer N, Saueressig F and Zanusso O 2015
{\it  Phys.\ Rev.} D {\bf 91} 025025

\end{thebibliography}
\end{document}